\documentclass[aps,prl,reprint,superscriptaddress,longbibliography]{revtex4-2}

\usepackage{amsmath,amssymb,bm}
\usepackage{graphicx}
\usepackage{xcolor}
\usepackage{hyperref}

\usepackage{amsmath,amssymb,amsfonts,bm,mathtools}
\usepackage{graphicx}
\usepackage{hyperref}
\usepackage{xcolor}

\hypersetup{colorlinks=true,citecolor=blue,linkcolor=blue,urlcolor=blue}

\newcommand{\Lg}{\mathbf{L}_{\textup{G}}}

\newcommand{\Lp}{\mathbf{L}_{\textup{P}}}
\newcommand{\aLp}{\boldsymbol{\mathcal{L}}_{\textup{P}}}

\newcommand{\kavg}{\left\langle k \right\rangle}
\newcommand{\vavg}{\left\langle v \right\rangle}

\newcommand{\gKM}{g_{\textup{KM}}}

\newcommand{\ipr}[1]{\mathrm{IPR}\left[#1\right]}

\newcommand{\mA}{\mathbf{A}}
\newcommand{\mD}{\mathbf{D}}

\newcommand{\mV}{\mathbf{V}}

\newcommand{\vx}{\mbox{\boldmath $x$}}

\newcommand{\vpsi}{\mbox{\boldmath $\psi$}}
\newcommand{\Cv}{C_{\textup{v}}}

\begin{document}

\title{Localization transitions of diffusion dynamics in physical networks}

\author{Jun Yamamoto}
\email{jun.j.yamamoto@gmail.com}
\affiliation{Department of Network and Data Science, Central European University, 1100 Vienna, Austria.}

\author{Ivan Bonamassa}
\email{ivan.bms.2011@gmail.com}
\affiliation{International Research Center for Complexity Science, Hangzhou International Innovation Institute (H3I) of Beihang University, Hangzhou, China.}

\author{M\'arton P\'osfai}
\email{posfaim@ceu.edu}
\affiliation{Department of Network and Data Science, Central European University, 1100 Vienna, Austria.}

\date{\today}

\begin{abstract}
Network diffusion underlies many transport phenomena, with Laplacian modes setting how information spreads and relaxes. In physical networks, however, connectivity alone is not enough: node volumes introduce local dwell times that regulate how flow is stored before being propagated. Here we show that physical heterogeneity reshapes topology-driven localization, with the degree-volume ratio emerging as the relevant disorder parameter. We solve an analytical model in which ratio detuning qualitatively reorganizes the Laplacian spectrum, and demonstrate in empirical networks how degree-volume correlations shift extremal eigenmodes away from the nodes selected by topology alone. Our results reveal a general feature-rich-driven mechanism for localization control, showing that physicality non-trivially reshapes the disorder landscape governing network dynamics.
\end{abstract}

\maketitle

Physical networks are not specified by connectivity alone~\cite{dehmamy2018structural, liu2021isotopy, blagojevic2025network}: nodes often represent extended components with heterogeneous physical properties, such as volume, mass, storage, or capacity, that can qualitatively change their function~\cite{pang2023geometric, ortiz2025unified, both2025scale, wytock2026irregular}.
While growing attention has recently been given to understanding how these physical properties shape network structure~\cite{posfai2024impact, glover2024measuring,blagojevic2024three, bonamassa2025logarithmic,piazza2025physical,meng2026surface}, the consequences of physicality on dynamics remain largely unexplored.
Graph operators, such as adjacency or Laplacian matrices, provide a bridge between structure and function~\cite{newman2003structure, boccaletti2006complex, mcgraw2008laplacian}, translating anomalous connectivity patterns into spectral properties that affect processes like diffusion, synchronization, or transport~\cite{arenas2008synchronization, goltsev2012localization, masuda2017random, ghavasieh2020statistical}.
Here, we study how physical and network structures interact to shape the Laplacian spectra and their eigenvectors in physical networks.

Focusing on diffusion dynamics, we show how network physicality leads to localization patterns and shifts in the Laplacian spectra that connectivity alone cannot predict. 
Building on a network-of-networks representation of physical networks~\cite{pete2024physical}, we find that physical Laplacians are governed by an interplay between degree and volume heterogeneities, with the degree-volume ratio emerging as the natural disorder variable. 
In particular, we demonstrate a counterintuitive effect: endowing the nodes of an abstract network with additional details, such as temporal activity, geographic location or other meta-data, is often expected to add heterogeneity to the system.
Here, we find that physical realization can instead cancel topological disorder or even invert the identity of spectral defects when node volumes grow sub- or super-linearly with node degree. 
As a result, degree-volume correlations suppress, restore, or relocate localization centers away from the hubs selected by topology alone. 
We propose a minimal model that analytically confirms this mechanism, and we validate our predictions across several empirical systems. 
Our work establishes a bridge between structure and function in physical networks, highlighting a general principle for eigenmode localization based on attribute-connectivity correlations. 
This principle extends to arbitrary feature-rich networks~\cite{artime2021percolation, aliakbarisani2025feature}, revealing how scalar node attributes can reshape their dynamical response.
Here we focus on the physical mechanism and a minimal solvable model, while a companion paper gives the full multiscale derivation, interacting-defect and motif theory, ensemble and finite-size analyses, and empirical volume controls~\cite{yamamoto2026laplacian}.

\paragraph{Model and operators.}

We consider a connected simple network $G$ with $N$ nodes, adjacency matrix $\mA$, degree sequence $\{k_i\}$, and ordinary graph Laplacian $\Lg = \mD - \mA$ where $\mD=\mathrm{diag}(k_1,\dots,k_N)$. 
Each node $i \in V$ is assigned a positive volume $v_i \in \mathbb{R}^{+}$, stored in a diagonal matrix $\mV = \mathrm{diag}(v_1,\dots,v_N)$.
On this network, we consider a linear relaxation dynamics of the form
\begin{align}\label{eq:relaxation}
  \mV \dot{\vx} = - \Lg \vx
\end{align}
with $\vx\in\mathbb{R}^N$ a vector of node variables.
Such dynamics arise naturally in network diffusion problems~\cite{kondor2002diffusion,masuda2017random}, from transient solute transport in pore networks~\cite{bryntesson2002pore, xiong2016review} to thermal transport in lumped-parameter thermal networks~\cite{mellor1991lumped,boglietti2009evolution, cipollini2025introduction} and protein propagation in brain networks~\cite{luo2026temporal}.
Equation~\eqref{eq:relaxation} identifies the diffusive operator $\aLp = \mV^{-1}\Lg$; for spectral analysis, we use the corresponding symmetric form
\begin{align}
  \Lp = \mV^{-1/2}\Lg\mV^{-1/2},
\end{align}
which is related to $\aLp$ by similarity. 
The operators $\aLp$ and $\Lp$, also known as vertex-weighted Laplacians in graph theory~\cite{chung1996combinatorial}, are derived in the companion paper~\cite{yamamoto2026laplacian} for diffusion dynamics using a network-of-networks representation of physical networks under time-scale separation between internal and inter-node relaxation~\cite{pete2024physical}.
The operators $\aLp$ and $\Lp$ are isospectral, and their eigenmodes differ by a volume-dependent rescaling of component weights, $v_i^{-1/2}$, for each node $i$.
In what follows, we analyze $\Lp$ for simplicity, the results apply to $\aLp$ through the similarity transformation.

\paragraph{Spectral observables.}

Matrix $\Lp$ has eigenvalues $0 = \lambda_1 < \lambda_2 \le \cdots \le \lambda_N$ with orthonormal eigenvectors $\{\vpsi_{\mu}\}_{\mu=1}^N$.
The zero mode $\vpsi_{1}$ has its component weights distributed proportionally to the square root of the volumes of the nodes, so $\psi_{1,i} \propto \sqrt{v_i}$.
We focus on the Fiedler mode $\vpsi_2$ and leading mode $\vpsi_N$, associated with $\lambda_2$ and $\lambda_N$ respectively, which govern the slowest and fastest relaxation timescales of the dynamics.
To quantify their localization, we adopt the inverse participation ratio
\begin{align}
\ipr{\vpsi_{\mu}} = \sum_{i=1}^N \left|\psi_{\mu,i}\right|^4,
\end{align}
where $\sum_i \left|\psi_{\mu,i}\right|^2 = 1$. 
Extended modes are characterized by $\mathrm{IPR}=O(N^{-1})$, while $\mathrm{IPR}=O(1)$ whenever a mode is localized on a finite set of nodes.
Transforming the eigenmodes of $\Lp$ yields those of $\aLp$, which have a direct physical interpretation for example, in porous material pore volumes set local storage, throats mediate exchange, and localized extremal modes identify regions controlling slow equilibration or fast local relaxation.
In what follows, we study how the localization of these extremal modes is reorganized by volume heterogeneity and degree-volume correlations relative to the topology-only baseline.

\paragraph{Solvable defect model.}

\begin{figure}[!t]
  \centering
  \includegraphics[width=\linewidth]{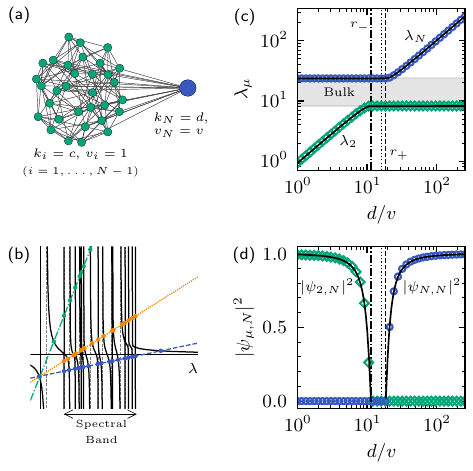}
  \caption{\label{fig:prl_fig1}
    (a)~The defect model.
    (b)~Graphical interpretation of Eq.~\eqref{eq:defect-extremal-eigenvalue}:
    Vertical lines are poles corresponding to the $N-2$ bulk eigenvalues and the trivial eigenvalue of the $c$-regular random graph before attaching the outlier node.
    The colored lines are the right-hand side for three values of $d/v$. 
    The dots mark the solutions, i.e., the $N$ new eigenvalues $\lambda_1,\lambda_2,\ldots,\lambda_N$ of the network after the perturbation by the outlier.
    If the rightmost perturbed eigenvalue $\lambda_N$ falls close to the edge of the bulk (orange), the location of the pole at the edge determines $\lambda_N$ and the other poles do not play a role, leading to a delocalized eigenvector similar to the eigenvectors of the unperturbed bulk.
    The localization transformation happens, when large detuning from $c$ pushes the rightmost solution outside of the bulk band (blue) and $\lambda_N$ is affected by the bulk through its Stieltjes transform, not an individual pole.
    The localization of the Fiedler mode happens similarly at the left edge of the bulk (green).
    (c)~Fiedler ($\lambda_2$) and leading ($\lambda_N$) eigenvalues of $\Lp$ and (d)~the corresponding outlier components $|\psi_2^{\mathrm{o}}|^2$ and $|\psi_N^{\mathrm{o}}|^2$ vs.\ outlier ratio $d/v$.
    The black lines show the theoretical predictions of Eqs.~\eqref{eq:defect-extremal-eigenvalue} and \eqref{eq:defect-eigvec-weight}, and the symbols are the results from direct diagonalization. 
    The shaded region in (c) indicates the bulk band $[c^\prime-2\sqrt{c-1},\, c^\prime+2\sqrt{c-1}]$.
    In all panels, $N=10^5$, $c=2^4$, and $d=2^7$, and the results are averaged over 100 realizations.
  }
\end{figure}

To isolate the mechanism by which physical node heterogeneity reorganizes extremal-mode localization, we first consider a solvable defect model. 
To construct our model, we start from a $c$-regular random graph of $N-1$ nodes forming a homogeneous bulk, and we assign unit volume to each node. 
Because all bulk nodes carry unit volume, the eigenvalues of $\Lp$ of the unperturbed bulk reduce to the eigenvalues of the regular Laplacian $\Lg$, which for $c$-regular random graphs follow the Kesten-McKay distribution with all corresponding eigenvectors delocalized with high probability~\cite{mckay1981expected,bauerschmidt2019local,huang2024spectrum}.
We now perturb the network by attaching a single outlier node of volume $v$ to the $d$ bulk nodes selected uniformly at random, where $d$ is the degree of the outlier (Fig.~\ref{fig:prl_fig1}(a)).
The bulk is therefore characterized by the bulk-mean degree-volume ratio, $c^\prime=c+d/(N-1)$, whereas the outlier carries the anomalous ratio $d/v$.
This model reduces to the standard degree-disorder setting when $v=1$ and to a pure volume-disorder setting when $d=c$, showing that degree heterogeneity and volume heterogeneity enter on equal footing by detuning $d/v$ from the bulk value $c^\prime$. 
As we show below, sufficiently strong detuning causes an extremal eigenvalue to split from the bulk spectral band (Fig.~\ref{fig:prl_fig1}(b)) and the corresponding eigenmode to localize on the outlier.

Following the approach of Ref.~\cite{martin2014localization}, we derive a self-consistent equation for the eigenvalues $\lambda_\mu$ of $\Lp$ (see~\cite{yamamoto2026laplacian}).
For the extremal eigenvalues, $\lambda_2$ and $\lambda_N$, lying outside the bulk band $[c^\prime-2\sqrt{c-1},\,c^\prime+2\sqrt{c-1}]$, the resolvent can be evaluated in the thermodynamic limit using the Stieltjes transform, $g_{\rm KM}(z;c)$, of the Kesten--McKay distribution, yielding
\begin{align}
  \frac{d^2}{(N-1)\lambda_{\mu}-d} - d \, \gKM(c^\prime - \lambda_{\mu}; c)
  = v \lambda_{\mu} - d. \label{eq:defect-extremal-eigenvalue}
\end{align}
Similarly, the squared weights of the outlier component can be written in terms of the derivative $\gKM^\prime=\partial \gKM(z; c)/\partial z$ of the Stieltjes transform as
\begin{align}
  \left|\psi_{\mu,N}\right|^2
  = \left[ 1 - \frac{d}{v}\,\gKM'(c^\prime - \lambda_{\mu};c) + \frac{(N-1)d^2}{v \left[(N-1)\lambda_{\mu}-d\right]^2} \right]^{-1},
  \label{eq:defect-eigvec-weight}
\end{align}
for $|\lambda_\mu-c^\prime|>2\sqrt{c-1}$. 
These expressions show that the bulk supports a spectral band centered around $c^\prime$, while a sufficiently anomalous outlier ratio $r \coloneq d/v$ detaches an extremal eigenvalue. More precisely, there exist critical values $r_{\mathrm{c}}^-<c^\prime<r_{\mathrm{c}}^+$ such that the Fiedler and leading modes localize on the outlier node for $r<r_{\mathrm{c}}^-$ and $r>r_{\mathrm{c}}^+$, respectively. As shown in Fig.~\ref{fig:prl_fig1}(c,d), simulations closely match the predictions of Eqs.~\eqref{eq:defect-extremal-eigenvalue} and \eqref{eq:defect-eigvec-weight}. Across these thresholds, the outlier weight continuously grows from zero to an $O(1)$ value, confirming that localization is controlled by the ratio $r=d/v$ rather than degree or volume alone.

\paragraph{Random graph validation.}

\begin{figure}[t]
  \centering
  \includegraphics[width=\columnwidth]{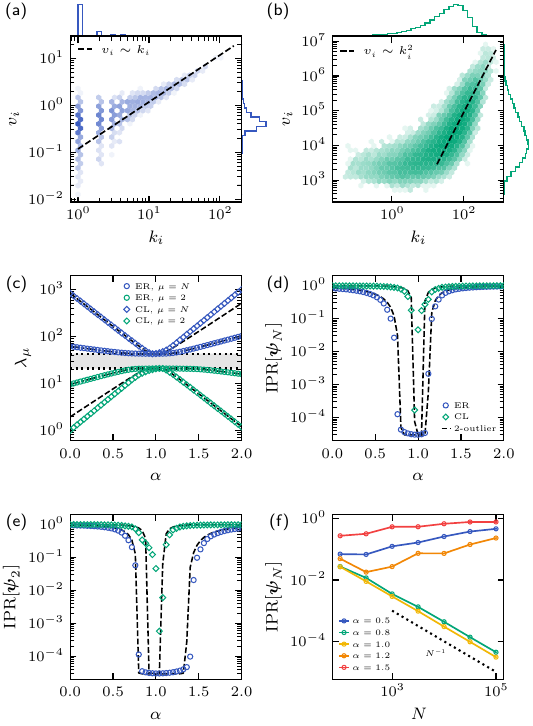}
  \caption{\label{fig:prl_fig2}
    (a, b)~Empirical degree-volume correlations in a vascular network (human liver) and a pore network (Bentheimer sandstone). Notice the power-law correlation in (a) and the constant offset $v_0$ in (a,b).
    (c)~Largest and smallest non-zero eigenvalues, (d,e)~the inverse participation ratios of the leading (d) and Fiedler (e) eigenvectors as a function of the degree-volume correlation exponent $\alpha$.
    For each realization of network models with $N=10^5$ nodes and mean degree $\langle k \rangle = 32$, node volumes are assigned deterministically by Eq.~\eqref{eq:volume-assignment} for each node $i$.
    The markers indicate the inverse participation ratios for node-weighted Chung--Lu model with power-law degree distribution $P(k) \sim k^{-2.5}$ (CL, blue circles) and Erd\H{o}s--R\'enyi random graphs (ER, green squares). 
    The dashed lines indicate the IPRs for the extended analytical model with 2 outlier nodes with the same mean degree $c=\langle k \rangle=32$, minimum degree $d_1 = \mathbb{E}[\min_i (k_i)]$ and maximum degree $d_2 = \mathbb{E}[\max_i (k_i)]$ as the random network ensembles.
    (f)~Finite-size scaling of the IPR of the leading mode in ER graphs with $\kavg=32$.
  }
\end{figure}

The outlier model above identifies the local degree-volume ratio, $k_i/v_i$, as the natural control variable for extremal-mode localization in the node-weighted Laplacian $\Lp$. To test whether this mechanism extends to more realistic settings with multiple degree outliers, we consider synthetic random graphs in which node volumes are assigned via the relation
\begin{align}
  v_i = \Cv k_i^\alpha + v_0, \qquad i = 1,2, \ldots, N \label{eq:volume-assignment}
\end{align}
where $\alpha,\,v_0\ge 0$ are parameters and $\Cv$ is fixed so that $\vavg = \kavg$. 
The latter choice ensures that the diagonal entries of $\Lp$ are of order $1$ and enables a meaningful comparison across networks of different mean degree.
The power-law form in Eq.~\eqref{eq:volume-assignment} reflects the empirical observation (Fig.~\ref{fig:prl_fig2}(a,b)) that degree and volume are positively correlated in physical networks~\cite{pete2024physical, piazza2025physical}, and defines a family of node-weighted Laplacians, $\Lp$, controlled by the degree-volume correlation exponent $\alpha$.
Notably, for $v_0=0$, the cases $\alpha=0$ and $\alpha=1$ recover, respectively, the ordinary and the degree-normalized Laplacians, so that $\alpha$ continuously interpolates between these well-known limits. 
Hence, since $v_0>0$ leaves the results qualitatively unchanged~\cite{yamamoto2026laplacian}, we set $v_0=0$ below.

Panels (c--e) in Fig.~\ref{fig:prl_fig2} show that varying $\alpha$ strongly reorganizes the localization of the extremal modes, as it modulates the ratio $r_i \sim k_{i}^{1-\alpha}$.
In Erd\H{o}s--R\'enyi (ER) random graphs, the Fiedler (leading) mode $\vpsi_{2}$ ($\vpsi_{N}$) is localized on small-ratio (large-ratio) nodes for $\alpha<1$, becomes extended near $\alpha=1$, and localizes again on small-ratio (large-ratio) nodes for $\alpha>1$.
Thus, the linear correlation $v_i \propto k_i$ marks a special point at which the degree-volume ratio is homogenized and extremal localization is strongly suppressed.
The behavior of $\ipr{\vpsi_{2}}$ and $\ipr{\vpsi_{N}}$, together with the finite-size analysis reported in Fig.~\ref{fig:prl_fig2}(f), indicates that this reorganization is sharp and that this behavior is not just a consequence of finite-size effects (see \cite{yamamoto2026laplacian} for further details). 
Chung--Lu (CL) scale-free random graphs display the same qualitative structure, though with narrower delocalization near $\alpha=1$. 
In both network ensembles, the localization centers are selected by anomalous degree-volume ratio rather than by degree alone. 
This behavior is reproduced approximately by a two-defect extension of the solvable model (dashed line in Fig.~\ref{fig:prl_fig2}c-e) where the bulk degree is set to the mean degree of the random graph ensembles and the two outliers are set to the expected minimum and maximum degrees of the ensembles~\footnote{Details about the two-defect model and multi-defect models are given in Ref.~\cite{yamamoto2026laplacian}.}.
The two-defect model is a crude approximation of the more realistic random graphs, yet it still captures the transitions with unexpected accuracy: near $\alpha=1$, degree heterogeneity is exactly counteracted by the node volumes and the details of the degree distribution become less important~\cite{yamamoto2026laplacian}. 

\begin{table*}[t]
\centering
\caption{
Empirical physical networks used in the analysis.
Network size $N$, mean (weighted) degree $\langle k \rangle$, and Spearman correlation coefficient $\varrho(k, v)$ between degrees and volumes in empirical physical networks.
Network construction details are presented in~\cite{yamamoto2026laplacian}.
}
\label{tab:empirical_datasets}
\begin{ruledtabular}
  \begin{tabular}{llllllll}
    Dataset & System & $N$ & $\langle k\rangle$ & $\varrho(k, v)$ & Volume definition & Weight definition & Ref. \\
    \hline
    Fruit fly hemibrain & Neuronal & $21{,}662$ & $3.986$ & $0.718$ & neuron volume & synapse count & \cite{scheffer2020connectome} \\
    Fruit fly full brain & Neuronal & $138{,}104$ & $3.234$ & $0.748$ & neuron volume & synapse count & \cite{dorkenwald2024neuronal,schlegel2024whole,zheng2018complete,yu2025new,buhmann2021automatic,heinrich2018synaptic} \\
    Fruit fly nerve cords & Neuronal & $101{,}840$ & $5.920$ & $0.752$ & neuron volume & synapse count & \cite{takemura2024connectome,marin2024systematic,cheong2023transforming} \\
    Bentheimer sandstone & Porous & $46{,}888$ & $16.027$ & $0.918$ & pore volume & throat conductance & \cite{lucas2022micro} \\
    Berea sandstone & Porous & $92{,}743$ & $12.803$ & $0.867$ & pore volume & throat conductance & \cite{lucas2022micro} \\
    Bandera brown sandstone & Porous & $147{,}064$ & $10.360$ & $0.794$ & pore volume & throat conductance & \cite{lucas2022micro} \\
    Human liver (model) & Vascular & $24{,}000$ & $2.00$ & $0.519$ & vessel volume & --- & \cite{jessen2022rigorous} \\
    Danube river & River & $21{,}248$ & $2.00$ & $0.642$ & segment length & --- & \cite{lehner2013global} \\
    Amazon river & River & $424{,}583$ & $2.00$ & $0.509$ & segment length & --- & \cite{lehner2013global}
  \end{tabular}
  \end{ruledtabular}
\end{table*}

\begin{figure}[t]
    \centering
    \includegraphics[]{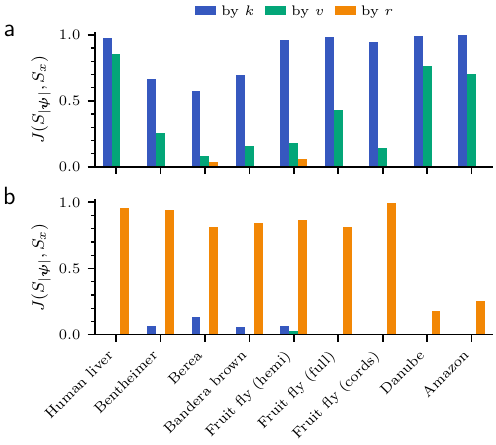}
    \caption{\label{fig:prl_fig3}
    Empirical validation in physical networks. 
    Overlap between the set $S_{|\psi|}$ of nodes with the largest weight in the leading 100 eigenvectors of $\Lg$ (a) and $\Lp$ (b) and the sets $S_{k}$, $S_{v}$, $S_{k/v}$ of nodes with the 100 largest degree, volume, and degree-volume ratio, respectively, quantified by the Jaccard index $J(S_{|\psi|}, S_x) = |S_{|\psi|} \cap S_x| / |S_{|\psi|} \cup S_x|$ ($x \in \{k,v,k/v\}$), which ranges from $0$ (no overlap) to $1$ (perfect overlap).
    }
\end{figure}

\paragraph{Empirical physical networks.}

To complete the analysis, we test the robustness of the ratio-based localization mechanism on empirical physical networks. 
For each dataset in Table~\ref{tab:empirical_datasets} and each operator $\Lg$ and $\Lp$, we identify the set $S_{|\psi|}$ of nodes carrying the largest eigenvector weights across the 100 largest-eigenvalue modes, and compare it against the sets $S_k$, $S_v$ and $S_{k/v}$ of nodes with the largest degree, volume, and degree-volume ratio, respectively, using the Jaccard index (see Ref.~\cite{yamamoto2026laplacian} for further details). 
For the ordinary Laplacian, $\Lg$ (Fig.~\ref{fig:prl_fig3}), we find that $S_{|\psi|}$ overlaps most strongly with $S_k$, consistent with the spectral graph theoretical picture in which localization is governed by degree heterogeneity. 
The positive degree-volume correlation in these networks also produces a secondary overlap with $S_v$, though weaker than that with $S_k$. 
For the node-weighted Laplacian $\Lp$, on the other hand, the dominant overlap shifts clearly to $S_{k/v}$, indicating that the large-weight components concentrate on nodes with the highest degree-volume ratio, confirming that $k_i/v_i$ is the relevant disorder variable for localization phenomena in physical networks.

\paragraph{Discussion.}

We showed that physical node heterogeneity does more than perturb the localization pattern of the ordinary graph Laplacian: it changes the notion of structural defects~\cite{biroli1999single, monasson1999diffusion}. 
In abstract networks, localization is driven by anomalous connectivity, because a node with degree far from the typical degree of its surroundings produces an outlying diagonal contribution to the local eigenvalue problem~\cite{mcgraw2008laplacian,martin2014localization,pastor2016distinct,hata2017localization}. 
In the physical Laplacian, this local contribution is divided by the node volume, so the relevant spectral defects are no longer purely topological but normalized by the nodes' volumes. 
Physicality, therefore, reshapes the spectral landscape of a network, redefining which nodes act as localization centers.

This has immediate consequences for relaxation, synchronization, and diffusive processes on networks~\cite{motter2005network, gomez2011explosive}, indicating that the dynamically influential nodes are not the highest-degree hubs but those with anomalously large or small degree-volume ratio, $k_i/v_i$. 
The special case $v_i\propto k_i$ is also the point at which $\aLp=\mV^{-1}\Lg$ becomes proportional to the degree-normalized Laplacian $\mD^{-1}\Lg$, whose spectrum underlies the enhanced synchronizability mechanism of Motter, Zhou, and Kurths~\cite{motter2005network}. 
In the present setting, the same normalization suppresses localization by homogenizing the local ratio $k_i/v_i$.
Centrality, influence, or spectral entropy metrics~\cite{kitsak2010identification, morone2015influence, de2016spectral} must therefore be reassessed in physical networks, where physical features matter as much as connectivity. 

This principle naturally extends beyond node volumes: we expect ratio-controlled localization to be relevant wherever a scalar node attribute rescales local response ---typical, e.g.,\ in coarse-grained and multilayer networks~\cite{de2023more, villegas2023laplacian, villegas2025multi} or when studying the stability of network dynamics~\cite{meena2023emergent}--- and to inform the design of materials with tailored mechanical properties~\cite{ortiz2025unified} and physical learning systems where node capacities select the spectrally dominant units~\cite{stern2021supervised, stern2025physical}. 

Just as temporal heterogeneity can modify diffusion by coupling local waiting-time statistics to spectral relaxation~\cite{luo2026temporal}, and multiscale architecture can be exploited to design scale-rich metamaterials~\cite{both2025scale, wytock2026irregular}, our results suggest that physical heterogeneity can qualitatively alter network dynamics by rescaling the graph operators. 
This points toward a theory of dynamics on physical networks in which connectivity, geometry, and physical node attributes jointly determine collective function.

{\em Acknowledgments}---
M.P., I.B., and J.Y. acknowledge support from ERC grant No.~810115-DYNASNET. 
J.Y. also acknowledges support from the AccelNet-MultiNet program, funded by the National Science Foundation under Awards No.~1927425 and No.~1927418.

{\em Data availability}---
The data that support the findings of this article are openly available~\cite{scheffer2020connectome,dorkenwald2024neuronal,schlegel2024whole,takemura2024connectome,lucas2022micro,jessen2022rigorous,lehner2013global}.

\bibliographystyle{apsrev4-2}
\bibliography{references.bib}

\end{document}